\documentclass{ws-ijmpd}
\usepackage[super,compress]{cite}
\usepackage{enumerate}
\begin{document}

\markboth{A.~Oliveros, Hern\'an E.~Noriega}
{{Constant-roll inflation driven by a scalar field}}

%%%%%%%%%%%%%%%%%%%%% Publisher's Area please ignore %%%%%%%%%%%%%%%
%
\catchline{}{}{}{}{}
%
%%%%%%%%%%%%%%%%%%%%%%%%%%%%%%%%%%%%%%%%%%%%%%%%%%%%%%%%%%%%%%%%%%%%

\title{Constant-roll inflation driven by a scalar field\\
with non-minimal derivative coupling}

\author{\footnotesize A.~Oliveros$^{*}$,  Hern\'an E.~Noriega$^{\dagger}$}

\address{Programa de F\'isica, Universidad del Atl\'antico, \\ Carrera 30 N\'umero 8-49 Puerto Colombia-Atl\'antico, Colombia\\
$^{*}$alexanderoliveros@mail.uniatlantico.edu.co\\
$^{\dagger}$henoriega@mail.uniatlantico.edu.co}

\maketitle

%\begin{history}
%\received{Day Month Year}
%\revised{Day Month Year}
%\end{history}

\begin{abstract}
In this work,  we study constant-roll inflation driven by a scalar field with non-minimal derivative coupling to gravity, via the Einstein tensor.   This model contains  a  free parameter, $\eta$, which quantifies the non-minimal derivative coupling and a parameter $\alpha$ which characterize the constant-roll condition. In this scenario,  using the Hamilton-Jacobi-like formalism, an ansatz for the Hubble parameter (as a function of the scalar field) and some restrictions on the model parameters, we found  new exact solutions for the inflaton potential which include power-law, de Sitter, quadratic hilltop  and natural inflation, among others. Additionally, a phase space analysis was performed and  it is shown that the  exact solutions associated to natural inflation and a ``$\cosh$-type'' potential,  are attractors. 
\end{abstract}


\ccode{PACS numbers: 98.80.Cq, 98.80.-k, 04.50.Kd}

%\tableofcontents

\section{Introduction}\label{sec_intro}

Inflation is a stage of the  early universe in which it exhibits a quasi-de Sitter phase  during a very short time ($\sim 10^{-33}\,s$) and at very high energy scales ($\sim 10^{16}\,eV$)  after the Big Bang. The inflationary paradigm  is currently considered as a part of the standard modern cosmology. This paradigm was introduced  in the early 1980's to resolve some problems that the Hot Big Bang model of the universe can't explain (e.g., the horizon, the flatness, and the monopole problems, among others) \cite{alexei1, guth, albrecht, linde}.  On the other hand, inflation makes predictions about  properties of the current universe which have been confirmed by numerous cosmological and astrophysical  observations (e.g., the temperature fluctuations in CMB spectrum \cite{hu, soda, wands}, the existence of large scale structures \cite{chibisov, alexei2, hawking, guth2}, and the nearly scale invariant primordial power spectrum \cite{alexei3, lyth, lidsey}).\\

\noindent The simplest scenario to explain the dynamics of inflation consists of  introducing a single scalar field (dubbed inflaton) minimally  coupled to gravity and with a nearly flat potential, namely, the potential energy of the field dominates over their kinetic term. Under this condition, it has the so-called slow-roll inflation (for more details about this topic see  Refs.~\citen{linde2} and \citen{liddle1}). On the other hand, it is also possible to consider the inflaton non-minimally coupled to gravity via the Ricci scalar, the Ricci tensor, the Gauss-Bonnet invariant, etc, which have been widely studied in the literature in distinct scenarios. All these theories (with minimal or non-minimal couplings to gravity) are known as scalar-tensor theories of gravitation (see e.g. Ref.~\citen{clifton} for a review). In this context, the most general scalar-tensor theory that produces second-order equations of motion was found by Horndeski \cite{horndeski}. A subclass of the Horndeski models, are theories in which the non-minimal derivative coupling (NMDC) with the Einstein tensor is taken into account.  But, at present, these theories  are not good candidates to explain dark energy (DE), since the detection of an electromagnetic counterpart (GRB 170817A) to the gravitational wave signal (GW170817) from the merger of two neutron stars, showed that the speed of gravitational waves (GW) $c_t$ is the same as the speed of light, within deviations of order $10^{-15}$ for the redshift $z<0.009$ (see  Refs.~\citen{abbot1} and \citen{abbot2}). Contrary  to this result, the theories with NMDC predict a variable GW speed at low redshift. In general, the above imposes stringent constraints on dark energy models  constructed in the framework of scalar-tensor and vector-tensor theories. Nevertheless, this restriction does not apply to high redshift values, thus, these scalar-tensor theories with NMDC could be used in an inflationary context (for a recent review about this topic see Ref.~\citen{kase1}).
In the literature, there are several  works on inflation with a NMDC. \cite{granda, shinji, gao1, gao2, saridakis} Usually, these models have been studied in the context of slow-roll inflation (see e.g. Refs.~\citen{myung}
and \citen{burin} for more details), but in the last years,  a new route has been considered in the literature instead of slow-roll inflation. In this new scenario, the slow-roll inflation is replaced by the more general, constant-roll condition. In these models the scalar field $\phi$ is assumed to satisfy the constant-roll condition $\ddot{\phi}=-(3+\alpha)H\dot{\phi}$, where $H$ is the Hubble parameter and $\alpha$ is an arbitrary constant. one can see that the usual slow-roll inflation occurs if $\alpha=-3$ while the ``ultra-slow-roll'' case corresponds to $\alpha=0$.  Constant-roll inflation was originally introduced in Ref.~\citen{martin} and recently it attracted a lot of interest \cite{motohashi1, motohashi2, alexei4,
odintsov1, odintsov2, odintsov3, pieroni, odintsov4, ito, karam, yi, gao3, morse}.\\

\noindent In this work we use the constant-roll condition in a model of inflation with NMDC to the Einstein tensor, we also use the Hamilton-Jacobi-like formalism, an ansatz for the Hubble parameter (as a function of the scalar field) and  some restrictions  on the model parameters. From the above, we find  new exact solutions for the inflaton potential.\\

\noindent  This paper it is organized as follows: in section \ref{sec_model} we introduce the scalar-tensor model of inflation with non-minimal derivative coupling to gravity, via the Einstein tensor, and the  corresponding field equations are obtained. In section \ref{sec_cosmo}, we consider a flat FRW  universe and a homogeneous scalar  field, and from these considerations,  general expressions for the equation of motion and the  total energy momentum tensor are obtained. In section \ref{sec_cosmo}  we consider the constant-roll condition and we use the Hamilton-Jacobi-like formalism, an ansatz for the Hubble parameter (as a function of the scalar field) and some restrictions on the model parameters, to find  new exact solutions for the inflaton potential.  In section \ref{sec4} we realize a phase space analysis, and some conclusions are exposed in section \ref{sec5}. 

\section{The model}\label{sec_model} 
\noindent  The action for the scalar field with the kinetic term non-minimally coupled to Einstein tensor is

\begin{equation}\label{eq1}
S=\int { { d }^{ 4 }x\sqrt { -g } \left(\frac { { M }_{ pl }^{ 2 } }{ 2 } R-\frac { 1 }{ 2 } { g }^{ \mu \nu  }{ \partial  }_{ \mu  }\phi { \partial  }_{ \nu  }\phi -V(\phi )+\eta\,{ G }_{ \mu \nu  }{ \partial  }^{ \mu  }\phi { \partial  }^{ \nu  }\phi  \right)  },
\end{equation}

\noindent where $M^2_{\text{pl}}=(8\pi G)^{-1}$ is the reduced Planck mass, $ g _{ \mu \nu  }$ is the metric tensor, $G_{ \mu \nu  }={ R }_{ \mu \nu  }-\frac { 1 }{ 2 } { g }_{ \mu \nu  }R$ is the Einstein tensor, $\eta$ is a coupling constant with dimensions $ M ^{ -2 }$ and $V(\phi)$ is a potential term.\\

\noindent Models with a NMDC similar to Eq. (\ref{eq1}) have been widely studied in the literature in a cosmological and astrophysical context
\cite{amendola, capozziello, daniel, saridakis2, burin2}. In cosmology, usually, they have been used to explain inflation and dark energy problems. However, recently these models have been discarded in the context of dark energy, since they predict a variable GW speed at low redshift (which is contrary to the gravitational waves measurements lately realized \cite{abbot1, abbot2}). In the inflationary context, there are several  works on inflation with a NMDC, for example,  in Ref.~\refcite{shinji},  the author studied observational constraints on a number of representative inflationary models with a field derivative coupling to the Einstein tensor. In Refs.~\refcite{gao1} and \refcite{gao2} the authors derive the general formulae for the the scalar and tensor spectral tilts to the second order for the inflationary models with NMDC taking into account high friction limit and without it. In Ref.~\refcite{myung}, the authors investigate the slow-roll inflation in the NMDC model with exponential, quadratic, and quartic potentials. Finally, in Ref.~\refcite{saridakis}  the authors investigate inflation with NMDC through the Hamilton-Jacobi formalism.\\

\noindent The variation of the action, Eq.~(\ref{eq1}), with respect to the metric tensor $g_{\mu\nu}$ gives the field equations
\begin{equation}\label{eq2}
{ R }_{ \mu \nu  }-\frac { 1 }{ 2 } { g }_{ \mu \nu  }R=\frac{1}{{ M }_{ pl }^{ 2 }}{ T }_{ \mu \nu  }, \hspace{0.8cm} { T }_{ \mu \nu}={ T }_{ \mu \nu  }^{ (\phi ) }+\eta \hspace{0.04cm} { \Theta  }_{ \mu \nu  }, 
\end{equation}
where ${ T }_{ \mu \nu  }^{ (\phi ) }$ and ${ \Theta  }_{ \mu \nu  }$ are given by
\begin{equation}\label{eq3}
{ { T }_{ \mu \nu  }^{ (\phi ) }}= {\nabla}_{ \mu  }\phi {\nabla }_{ \nu  }\phi -{ g }_{ \mu \nu  }\Big({ \nabla  }^{ \alpha  }\phi { \nabla  }_{ \alpha  }\phi +V(\phi )\Big),
\end{equation}

\begin{align} \label{eq4}
{ \Theta  }_{ \mu \nu  }= & \hspace{0.06cm} { G }_{ \mu \nu  }{ \nabla  }^{ \alpha  }\phi { \nabla  }_{ \alpha  }\phi +{ g }_{ \mu \nu  }\Big({ R }^{ \alpha \beta  }{ \nabla  }_{ \alpha  }\phi { \nabla  }_{ \beta  }\phi+\Box \left( { \nabla  }^{ \alpha  }\phi { \nabla  }_{ \alpha  }\phi  \right) -{ \nabla  }^{ \alpha  }{ \nabla  }^{ \beta  }\left( { \nabla  }_{ \alpha  }\phi { \nabla  }_{ \beta  }\phi  \right) \Big)+R{ \nabla  }_{ \mu  }\phi { \nabla  }_{ \nu  }\phi \nonumber \\ &+2{ \nabla  }_{ \alpha  }{ \nabla  }_{ (\mu  }\left( { \nabla  }_{ \nu ) }\phi { \nabla  }^{ \alpha  }\phi  \right)-4{ R }_{ \alpha (\mu  }{ \nabla  }_{ \nu ) }\phi { \nabla  }^{ \alpha  }\phi -{ \nabla  }_{ \nu  }{ \nabla  }_{ \mu  }\left( { \nabla  }^{ \alpha  }\phi { \nabla  }_{ \alpha  }\phi  \right) -\square \left( { \nabla  }_{ \mu  }\phi { \nabla  }_{ \nu  }\phi  \right).
\end{align}

\noindent On the other hand, the variation of the action with respect to the scalar field, gives the equation of  motion 
\begin{equation}\label{eq6}
-\square\left( \phi  \right) +\frac { \partial V }{ \partial \phi  } +2\eta { \nabla  }_{ \mu  }\left( { R }^{ \mu \nu  }{ \partial  }_{ \nu  }\phi  \right) -\eta { \nabla  }_{ \mu  }\left( { R }{ \partial  }^{ \mu  }\phi  \right)=0.
\end{equation}

\section{Constant-roll inflation with NMDC}\label{sec_cosmo}
\noindent In this section we use the constant-roll condition in the background equations obtained in the previous section. To this aim, the first step is to consider a flat, homogeneous and isotropic universe whose metric  is given by Friedmann-Robertson-Walker (FRW) metric:
\begin{equation}\label{eq7}
ds^2=-dt^2+a(t)^2 \delta_{ij}dx^idx^j,
\end{equation}
where $a(t)$ is the scale factor. Using in Eqs. (\ref{eq2}) and (\ref{eq6}), the FRW metric and considering that the scalar field is homogeneous ($\phi\equiv \phi(t)$), we obtain the Friedmann equations
\begin{equation}\label{eq8}
3{ M }_{ pl }^{ 2 }{ H }^{ 2 }=\frac { 1 }{ 2 } { \dot { \phi  }  }^{ 2 }+V\left( \phi  \right) +9\eta { H }^{ 2 }{ \dot { \phi  }  }^{ 2 },
\end{equation}

\begin{equation}\label{eq9}
-2{ M }_{ pl }^{ 2 }\dot { H } ={ \dot { \phi  }  }^{ 2 }+6\eta { H }^{ 2 }{ \dot { \phi  }  }^{ 2 }-2\eta \dot { H } { \dot { \phi  }  }^{ 2 }-4\eta H\dot { \phi  } \ddot { \phi  },
\end{equation}
and the scalar field equation
\begin{equation}\label{eq10}
\ddot { \phi  } +3H\dot { \phi  } +\frac { dV }{ d\phi  } +12\eta H\dot { H } \dot { \phi  } +6\eta { H }^{ 2 }\ddot { \phi  } +18\eta { H }^{ 3 }\dot { \phi  } =0,
\end{equation}
where a dot denotes differentiation with respect to cosmic time $t$.\\

\noindent Usually, to solve the system of equations (\ref{eq8})-(\ref{eq10}) in an inflationary context,  it's a common practice to use the slow-roll parameters defined as
\begin{equation}\label{equ11}
\epsilon\equiv -\frac{\dot{H}}{H^2},
\end{equation}
\begin{equation}\label{equ12}
\delta\equiv -\frac{\ddot{\phi}}{H\dot{\phi}},
\end{equation}
which control the inflationary dynamics. Additionally, the  slow-roll conditions $|\epsilon|\ll 1$ and $|\delta|\ll 1$ are imposed.
In this work, let's consider that the second slow-roll condition is violated. In this sense, we use the constant-roll condition given by
\cite{motohashi1}
\begin{equation}\label{eq11}
\ddot { \phi  } =-\left( 3+\alpha  \right) H\dot { \phi  },
\end{equation}
where $\delta\equiv 3+\alpha$ and $\alpha$ an arbitrary constant. In order to solve the system of equations (\ref{eq8})-(\ref{eq10}) with the condition (\ref{eq11}), we consider the Hubble parameter as a function of the inflaton field $H = H(\phi)$ and use the Hamilton-Jacobi-like formalism \cite{motohashi1}.
Thereby, since $\dot{H} =\dot{\phi}\,{dH}/{d\phi}$, and using Eq. (\ref{eq11}), we can rewrite (\ref{eq9}) as
\begin{equation}\label{eq12}
-2{ M }_{ pl }^{ 2 }\frac { dH }{ d\phi  } =\dot { \phi  } \left( 1+6\eta { H }^{ 2 } \right) -2\eta \frac { dH }{ d\phi  } { \dot { \phi  }  }^{ 2 }+4\eta \left( \alpha +3 \right) { H }^{ 2 }\dot { \phi  },
\end{equation}
which can be solved with respect to $\dot{\phi}$,  so that
\begin{equation}\label{eq13}
\dot { \phi  } =\frac { 1+2\left( 9+2\alpha  \right) \eta { H }^{ 2 }\pm \sqrt { { \Big( 1+2\left( 9+2\alpha  \right) \eta { H }^{ 2 } \Big)  }^{ 2 }+16{ M }_{ pl }^{ 2 }\eta \left(\frac { dH }{ d\phi }\right)^{ 2 } }  }{ 4\eta \frac { dH }{ d\phi  } }.
\end{equation}
Notice that in  Eq. (\ref{eq13}) there are two possible solutions to $\dot{\phi}$. In this case, the right solution corresponds to the negative sign (since for $\eta=0$ the present model is reduced to the model analyzed in Ref.~\refcite{motohashi1}).\\

\noindent Differentiating Eq. (\ref{eq13}) with respect to $t$ and using Eq. (\ref{eq11}), we obtain
\begin{align}\label{eq14}
&12\left( 4+\alpha  \right) H+  \frac { \left[ -1-2\left( 9+2\alpha  \right) \eta { H }^{ 2 }+f(\phi)\right] \frac { d^2H }{ d\phi^2 } }{ \eta \left(\frac { dH }{ d\phi }\right)^{ 2 } } \nonumber \\ & -\frac { 4\left[ \left( 9+2\alpha  \right) H+2{ \left( 9+2\alpha  \right)  }^{ 2 }\eta { H }^{ 3 }+4{ M }_{ pl }^{ 2 }\frac { d^2H }{ d\phi^2 }\right]  }{f(\phi)} =0,
\end{align}

where, for simplicity, we have defined 
$f(\phi)\equiv\sqrt { { \left[ 1+2\left( 9+2\alpha  \right) \eta { H }^{ 2 } \right]  }^{ 2 }+16{ M }_{ pl }^{ 2 }\eta \left(\frac { dH }{ d\phi }\right)^{2} }$. It is evident that this non-linear differential equation for $H(\phi)$ is very complicated, and finding  exact solutions for it is a hard task. Nevertheless, we know that for $\eta=0$, Eq. (\ref{eq12}) is reduced to 
 \begin{equation}\label{eq15}
 -2{ M }_{ pl }^{ 2 }\frac { dH }{ d\phi  } =\dot { \phi  },
 \end{equation}
in this case,  Eq. (\ref{eq14}) becomes
\begin{equation}\label{eq16}
\frac { { d }^{ 2 }H }{ d{ \phi  }^{ 2 } } =\frac { 3+\alpha  }{ 2{ M }_{ pl }^{ 2 } } H,
\end{equation}
and the general solution of this equation is
\begin{equation}\label{eq17}
H\left( \phi  \right) ={ C }_{ 1 } \hspace{0.05cm} \textrm{exp}\left( \sqrt { \frac { 3+\alpha  }{ 2 }  } \frac { \phi  }{ { M }_{ pl } }  \right) +{ C }_{ 2 } \hspace{0.05cm} \textrm{exp}\left( -\sqrt { \frac { 3+\alpha  }{ 2 }  } \frac { \phi  }{ { M }_{ pl } }  \right).
\end{equation}
which  was found in Ref.~\refcite{motohashi1}. Inspired by this solution, the authors in Ref.~\refcite{ito} propose the following ansatz:
\begin{equation}\label{eq18}
H(\phi) ={ C }_{ 1 } \hspace{0.05cm} \textrm{exp}\left( \lambda \left(n\right) \sqrt { \frac { 3+\alpha  }{ 2 }  } \frac { \phi  }{ { M }_{ pl } }  \right) +{ C }_{ 2 }\hspace{0.05cm}\textrm{exp}\left( -\lambda \left(n\right) \sqrt { \frac { 3+\alpha  }{ 2 }  } \frac { \phi  }{ { M }_{ pl } }  \right).
\end{equation}
In a similar way, we will use this ansatz at the present model (changing $n$ by $\eta$).\\

\noindent Replacing the ansatz (\ref{eq18}) in Eq. (\ref{eq14}) and after some simplifications, we obtain the following system of algebraic equations:

\begin{equation}\label{eq19}
{ C }_{ 1 } \eta  \lambda (\alpha +3) (\alpha +4) (2 \alpha +9)=0,
\end{equation}

\begin{equation}\label{eq20}
{C}_{1} \eta  \lambda (\alpha +3) \Big[ 8 \alpha ^2+(4 \alpha +15)^2 \lambda ^2+66 \alpha -4 (\alpha +6) (2 \alpha +9)^2 {C}_{1} {C}_{2} \eta +135 \Big]=0,
\end{equation}

\begin{equation}\label{eq21}
\begin{aligned}
&{C}_{1} \eta  \lambda (\alpha +3) \Big[8 \alpha  \lambda ^2+\alpha+30 \lambda ^2+6 +656 \alpha ^3 {C}_{1}^{2}{C}_{2}^{2} \eta ^2+16 \alpha ^2 {C}_{1} {C}_{2} \eta  (549 {C}_{1} {C}_{2} \eta +16 \lambda ^2+6)\\ & +36 \alpha {C}_{1} {C}_{2} \eta(1089 {C}_{1} {C}_{2} \eta +56 \lambda ^2+24)+72 {C}_{1} {C}_{2} \eta  (810 {C}_{1} {C}_{2} \eta +55 \lambda ^2+27)\Big]=0,
\end{aligned}
\end{equation}

\begin{equation}\label{eq22}
\begin{aligned}
&{C}_{1} \eta  \lambda (\alpha +3) \Big[\Big(4 (2 \alpha +9) {C}_{1} {C}_{2} \eta +1\Big)^2 \Big(4 (7 \alpha +30) {C}_{1} {C}_{2} \eta +
1\Big)-\\
&\Big(4 (8 \alpha +33) {C}_{1} {C}_{2} \eta  \lambda +\lambda\Big)^2\Big]=0,
\end{aligned}
\end{equation}

\begin{equation}\label{eq23}
\begin{aligned}
&{C}_{1} {C}_{2} \eta  \lambda (\alpha +3) \Big[8 \alpha  \lambda ^2+\alpha+30 \lambda ^2+6 +608 \alpha ^3 {C}_{1}^{2} {C}_{2}^{2} \eta ^2+32 \alpha ^2 {C}_{1} {C}_{2} \eta (255 {C}_{1} {C}_{2} \eta +\\
&8 \lambda ^2+3) +72 \alpha  {C}_{1} {C}_{2} \eta  (507 {C}_{1} {C}_{2} \eta +28 \lambda ^2+12) +72 {C}_{1} {C}_{2} \eta (756 {C}_{1} {C}_{2} \eta +55 \lambda ^2+27)\Big]=0,
\end{aligned}
\end{equation}

\begin{equation}\label{equ23}
\begin{aligned}
&{C}_{1} {C}_{2} \eta  \lambda (\alpha +3) \Big[16 (2 \alpha +9)^2 (29 \alpha +126) {C}_{1}^{3} {C}_{2}^{3} \eta ^3 +4 {C}_{1} {C}_{2} \eta  (-2 (8 \alpha +33) \lambda ^2+\\
&11 \alpha +48)-\lambda ^2 +1+4 {C}_{1}^2 {C}_{2}^2 \eta ^2 \Big(3 (2 \alpha +9) (20 \alpha +87)-
(8 \alpha  (34 \alpha +279)+4581) \lambda ^2\Big)\Big]=0.
\end{aligned}
\end{equation}

\noindent The solutions of the system  (\ref{eq19})-(\ref{equ23}) are:
\begin{enumerate}[(a)]
\item $\{C_1=0\}$
\item $\{\alpha=-3\}$
\item $\{\lambda=0\}$
\item $\{C_2=0, \ \ \lambda=\pm 1, \ \ \alpha=-4\}$ 
\item $\{\lambda=\pm \sqrt{1+8C_1C_2\eta},\ \ \alpha=-4\}$
\end{enumerate}

\noindent  We now proceed to analyze the above solutions.
\subsection{Solution (a) $\{C_1=0\}$}
\begin{figure}[t]
\centerline{\includegraphics[width=1\linewidth,scale=1]{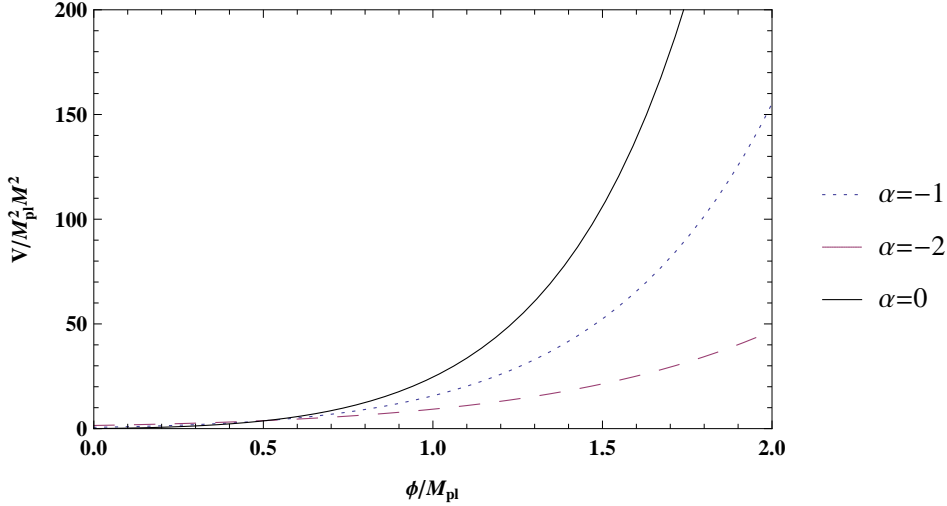}}
\caption{Potential versus scalar field, using $\gamma=1$, $\lambda=-1$ and  three different
values for $\alpha$:  $\alpha=-1$ (dotted line), $\alpha=-2$ (dashed line) and  $\alpha=0$ (solid line). \label{fig1}}
\end{figure}

\noindent In this case Eq. (\ref{eq18})  reduces to
\begin{equation}\label{eq24}
H(\phi)=Me^{-\lambda(\eta)\sqrt { \frac { 3+\alpha  }{ 2 }  } \frac { \phi  }{ { M }_{ pl } }},
\end{equation}
where the choice $C_2=M$ was made (the mass $M$ determines both the energy scale at which inflation occurs and the amplitude of the primordial fluctuations). Besides, in Eq. (\ref{eq24}),  $\lambda(\eta)$ and $\alpha$ are arbitrary quantities.
In this case, for $\alpha<-3$ and $\lambda(\eta)$ a real number, the solution (\ref{eq24}) has not  physical meaning. Conversely,
for  $\alpha>-3$ and $\lambda(\eta)$ a real  number, the solution (\ref{eq24}) is viable, and from Eq. (\ref{eq8}), we obtain
\begin{equation}\label{eq25}
\begin{aligned}
&V(\phi)=3 M^2 M_{\text{pl}}^2 e^{2 \beta\phi/M_{\text{pl}}}-\\
&\frac{M^2e^{-2 \beta  \phi/M_{\text{pl}} }\left(1+18 \gamma e^{2 \beta  \phi/M_{\text{pl}} }\right) \left[\sqrt{1+4 \gamma  e^{2 \beta\phi/M_{\text{pl}} }\left(1+4 \beta ^2M_{\text{pl}}^2+\gamma  e^{2 \beta\phi/M_{\text{pl}}}\right)}-1-2 \gamma  e^{2 \beta\phi/M_{\text{pl}} }\right]^2}{32 \beta^2 \gamma^2 },
\end{aligned}
\end{equation}
where $\gamma\equiv M^2\eta$ is a dimensionless parameter and $\beta=-\lambda(\eta)\sqrt{\frac{3+\alpha}{2}}$.  For suitable values of the model parameters, the potential (\ref{eq25}) represents a power-law-type inflation (see Fig.~\ref{fig1}). In addition, if $\eta\rightarrow 0$
and $\lambda=-1$, we have the power-law potential reported in Ref.~\refcite{motohashi1}, but it is ruled out by the authors (since it does not satisfy the observational constraints). In Ref.~\refcite{shinji} the author studied observational constraints on a number of representative inflationary models with NMDC, and in particular, he showed that exponential potentials (i.e., $V(\phi)=V_0e^{\beta\phi/M_{\text{pl}}}$, where $V_0$ and $\beta$ are constants) can be made compatible with the current observational data. However, in our case the potential (\ref{eq25}) is more complicate and it would be necessary to investigate the  most recent observational constraints on model parameters.\\

\subsection{ Solutions (b)  $\{\alpha=-3\}$ and (c) $\{\lambda=0\}$}
\noindent Continuing  with the analysis,  Eq. (\ref{eq18}) reduces to a constant and therefore these solutions represent the de Sitter universe.
\subsection{Solution (d) $\{C_2=0, \ \ \lambda=\pm 1, \ \ \alpha=-4\}$}
In this case  Eq. (\ref{eq18}) has not  physical meaning.
\subsection{Solution (e) $\{\lambda=\pm \sqrt{1+8C_1C_2\eta},\ \ \alpha=-4\}$}
\noindent This solution  is more interesting. In this case,
considering the positive sign in $\lambda$, Eq. (\ref{eq18}) takes the form
\begin{equation}\label{eq26}
H(\phi)=C_1e^{i\lambda(\eta)\frac{\sqrt{2}}{2}\frac{\phi}{M_{\text{pl}}}}+
C_2e^{-i\lambda(\eta)\frac{\sqrt{2}}{2}\frac{\phi}{M_{\text{pl}}}},
\end{equation}
where $\lambda(\eta)=\sqrt{1+8C_1C_2\eta}$. An identical result is obtained using the negative sign for $\lambda$.
Now, we consider the case $C_{1}=C_{2}={ M }/{ 2 }$ (a similar consideration was carried out in Refs.~\refcite{motohashi1} and \refcite{ito}), then (\ref{eq26}) becomes
\begin{equation}\label{eq27}
H(\phi) =M \cos {\left(\frac {\sqrt{2}}{2}\sqrt{1+2\gamma}\frac{\phi }{M_{\text{pl}}}\right)} ,
\end{equation}
where the restriction $1+2\gamma>0$ (or $\gamma>-\frac{1}{2}$) has been considered.\\ 

\noindent The potential $V(\phi)$ is obtained replacing  Eqs. (\ref{eq13}) and (\ref{eq26})  in 
Eq. (\ref{eq8}), so that
\begin{equation}\label{eq28}
\begin{aligned}
V(\phi)= &\frac{M^2M_{\text{pl}}^2}{4\left(1+2\gamma\right)}\Bigg[4+3\gamma+4(2+3\gamma)\cos{\left(\sqrt{2(1+2\gamma)}\frac{\phi}{M_{\text{pl}}}\right)}+\\
&9\gamma\cos{\left(2\sqrt{2(1+2\gamma)}\frac{\phi}{M_{\text{pl}}}\right)}\Bigg].
\end{aligned}
\end{equation}
\begin{figure}[t]
\centerline{\includegraphics[width=1\linewidth,scale=1]{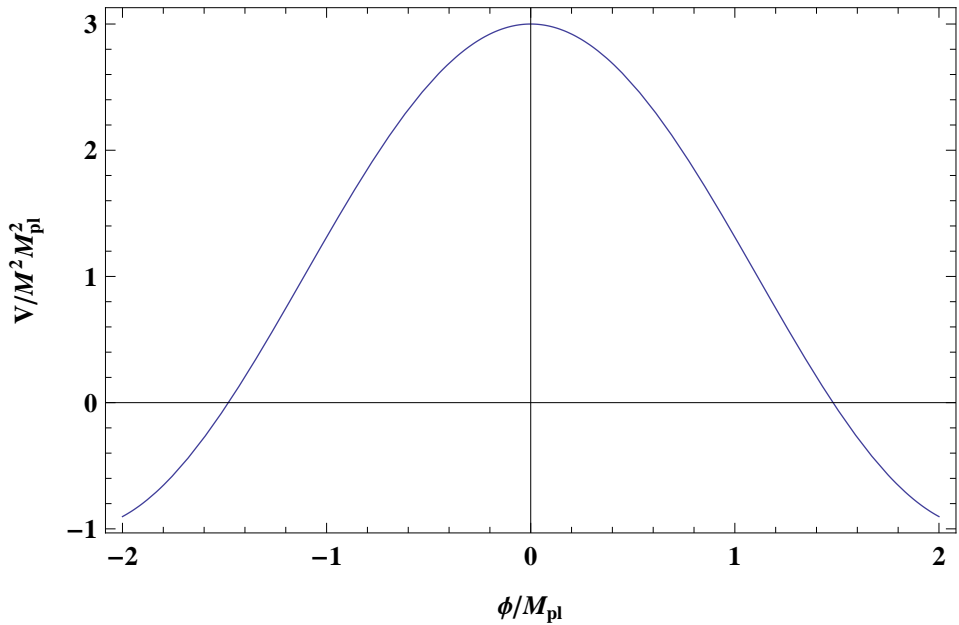}}
\caption{Potential versus scalar field, using $\gamma=10^{-9}$ in Eq. (\ref{eq28}). \label{fig2}}
\end{figure}
The potential (\ref{eq28})  represents  a general case of natural inflation (see Fig. \ref{fig2}). Considering $\phi$ near to origin, the potential (\ref{eq28}) reduces to  hilltop inflation \cite{lyth2}, namely
\begin{equation}\label{eq29}
V(\phi)\approx 3 M^2 M_{\text{pl}}^2-2 M^2\left(1+6 \gamma\right)\phi ^2,
\end{equation}
and to guarantee the hilltop inflation, $1+6\gamma>0$. Additionally, in Ref.~\refcite{motohashi1} the authors showed that a similar model to that given by Eq. (\ref{eq28}) is observationally viable.\\

\noindent In other words, replacing Eq. (\ref{eq27}) in Eq. (\ref{eq13}), we obtain 
\begin{equation}\label{eq30}
\phi(t)=\frac{2\sqrt{2}}{\sqrt{1+2\gamma}}M_{\text{pl}}\,\text{arctan}{\left(e^{Mt}\right)}.
\end{equation}
The integration constant that arises in (\ref{eq30}) has been removed, since  its contribution does not change the form
of the function $\phi(t)$. But, in general, $t$  must be replaced by $t+\frac{\sqrt{2(1+2\gamma)}}{2M M_{\text{pl}}}C$, where $C$ is the integration constant. Using Eq. (\ref{eq30}) in Eq. (\ref{eq27}), we get
\begin{equation}\label{eq31}
H(t)=-M\tanh(Mt),
\end{equation}
from the defnition $H=\dot{a}/a$, we have
\begin{equation}\label{eq32}
a(t)\propto\text{sech}(Mt).
\end{equation}
Additionally $\ddot{a}(t)>0$, so the model can describe an inflationary regimen.  Eqs. (\ref{eq30})-(\ref{eq32}) are similar (taking $\alpha=-4$) to those reported in Ref.~\refcite{motohashi1}. It is easy to check that if $Mt\rightarrow -\infty$ then $(\phi(t), \dot{\phi}(t))\rightarrow (0,0)$ and $H\rightarrow M$, which guarantees that these functions exhibit a good behavior under these conditions (i.e. are bounded functions). \\

\begin{figure}[t]
\centerline{\includegraphics[width=1\linewidth,scale=1]{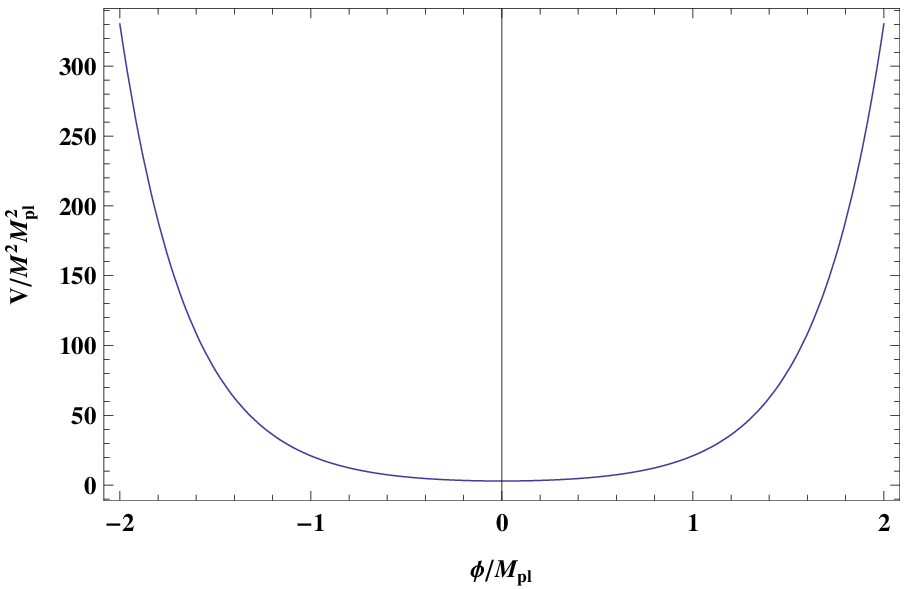}}
\caption{Potential versus scalar field, using $\gamma=-1$ in Eq. (\ref{eq34}). \label{fig3}}
\end{figure}

\noindent On the other hand, for $1+2\gamma<0$ (which implies that $\gamma<-\frac{1}{2}$), the ``$\cos$'' function
must be replaced by ``$\cosh$'' and $1+2\gamma$ by $|1+2\gamma|$ in Eq. (\ref{eq27}), namely
\begin{equation}\label{eq33}
H(\phi) =M \cosh {\left(\frac {\sqrt{2}}{2}\sqrt{|1+2\gamma|}\frac{\phi }{M_{\text{pl}}}\right)}.
\end{equation}
The potential $V(\phi)$ is
\begin{equation}\label{eq34}
\begin{aligned}
V(\phi)= &\frac{M^2M_{\text{pl}}^2}{4(1+2\gamma)}\Bigg[4+3\gamma + 4 (2+3\gamma) \cosh{\Bigg(\sqrt{2|1+2\gamma|}\frac{\phi }{M_{\text{pl}}}\Bigg)}\\
&+ 9\gamma\cosh {\Bigg(2\sqrt{2|1+2\gamma|}\frac{\phi }{M_{\text{pl}}}\Bigg)}\Bigg],
\end{aligned}
\end{equation}
and
\begin{equation}\label{eq35}
\phi(t)=\frac{2\sqrt{2}}{\sqrt{|1+2\gamma|}}M_{\text{pl}}\,\text{arctanh}{\left(e^{Mt}\right)},
\end{equation}
\begin{equation}\label{eq36}
H(t)=M\cosh[2\,\text{arctanh}\left(e^{Mt}\right)],
\end{equation}
\begin{equation}\label{eq37}
a(t)\propto\text{csch}(Mt).
\end{equation}
Also, in this case $\ddot{a}(t)>0$ and in general, $t\rightarrow t+\frac{\sqrt{2|1+2\gamma|}}{2M M_{\text{pl}}}C$. Again, if $Mt\rightarrow -\infty$ then $(\phi(t), \dot{\phi}(t))\rightarrow (0,0)$ and $H\rightarrow M$.
In  Fig. \ref{fig3} we can see the potential given by Eq. (\ref{eq34}).\\

\noindent Next, we consider the choice $C_1=M/2$ and $C_2=-M/2$ in Eq. (\ref{eq26}). In this case, for $1-2\gamma>0$ (or $\gamma<\frac{1}{2}$) the solution is not physical. For $1-2\gamma<0$ (or $\gamma>\frac{1}{2}$) and taking into account that
$\lambda=\pm \sqrt{|1-2\gamma|}$,  Eq. (\ref{eq26}) is reduced to
\begin{equation}\label{eq38}
H(\phi) =\mp M\sinh{\left(\frac {\sqrt{2}}{2}\sqrt{|1-2\gamma|}\frac{\phi }{M_{\text{pl}}}\right)}.
\end{equation}
The potential is
\begin{equation}\label{eq39}
\begin{aligned}
V(\phi)= &\frac{M^2M_{\text{pl}}^2}{4\gamma^2}\Bigg[6\gamma\Bigg(3-7\gamma+\gamma\,\text{cosh}\Bigg(\sqrt{2|1-2\gamma|}\frac{\phi }{M_{\text{pl}}}\Bigg)\Bigg)\\
&+(1-4\gamma(5-9\gamma))\text{sech}^2\Bigg(\frac{\sqrt{2|1-2\gamma|}}{2}\frac{\phi }{M_{\text{pl}}}\Bigg)\Bigg],
\end{aligned}
\end{equation}
which is valid for both signs of Eq. (\ref{eq38}). On the other hand, for the positive sign in (\ref{eq38}), we get
\begin{equation}\label{eq40}
\phi(t)=\mp\frac{\sqrt{2}}{\sqrt{|1-2\gamma|}}M_{\text{pl}}\,\text{arcsinh}{\left[ \pm Mt\left(1-\frac{1}{2\gamma}\right)\right]},
\end{equation}
and for the negative sign, we obtain
\begin{equation}\label{equ40}
\phi(t)=\pm\frac{\sqrt{2}}{\sqrt{|1-2\gamma|}}M_{\text{pl}}\,\text{arcsinh}{\left[ \pm Mt\left(1-\frac{1}{2\gamma}\right)\right]},
\end{equation}
and by last
\begin{equation}\label{eq41}
H(t)= M^2t\left(1-\frac{1}{2\gamma}\right),
\end{equation}
\begin{equation}\label{eq42}
a(t)\propto \exp{\left[\frac{M^2t^2}{2}\left(1-\frac{1}{2\gamma}\right)\right]}.
\end{equation}
which are valid for both signs.\\
 
\noindent Now, if $Mt\rightarrow \pm\infty$ then $(\phi(t), \dot{\phi}(t))\rightarrow (\pm\infty,0)$ and $H\rightarrow \pm\infty$. Therefore, the functions $\phi$ and $H$ do not present a good behavior under these conditions, and so, the solution  Eq. (\ref{eq38}) is not viable in this context.

\section{Analysis of the phase space}\label{sec4}
\begin{figure}[t]
\centerline{\includegraphics[width=1\linewidth,scale=1]{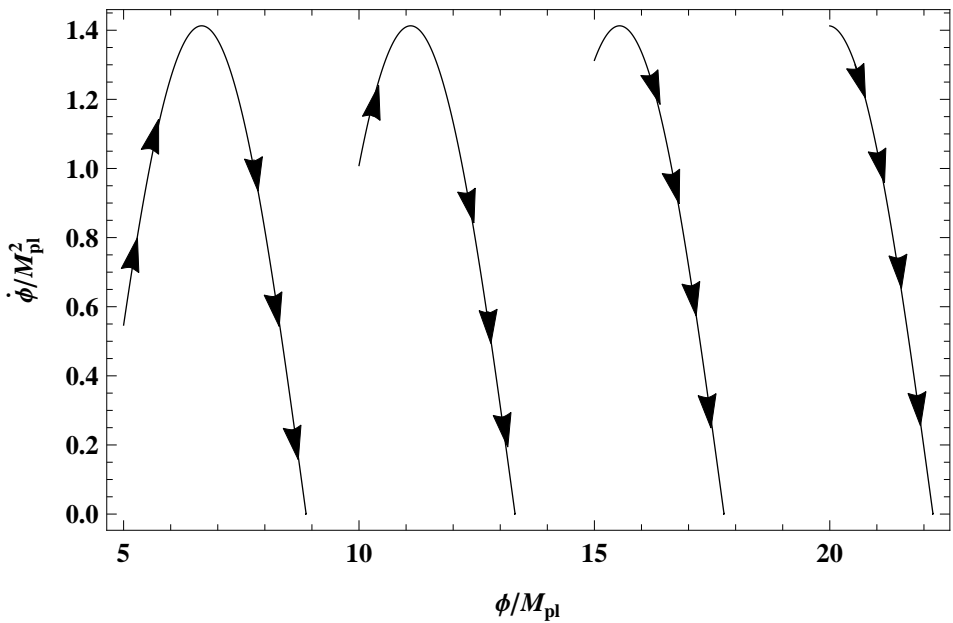}}
\caption{Phase space structure of the solution (\ref{eq27}).  In this numerical simulation we
have used:  from left to right $\frac{\phi_0}{M_{\text{pl}}}=5$, $\frac{\phi_0}{M_{\text{pl}}}=10$, $\frac{\phi_0}{M_{\text{pl}}}=15$,
$\frac{\phi_0}{M_{\text{pl}}}=20$ and  $\gamma=10^{-3}$. \label{fig4}}
\end{figure}

\begin{figure}[t]
\centerline{\includegraphics[width=1\linewidth,scale=1]{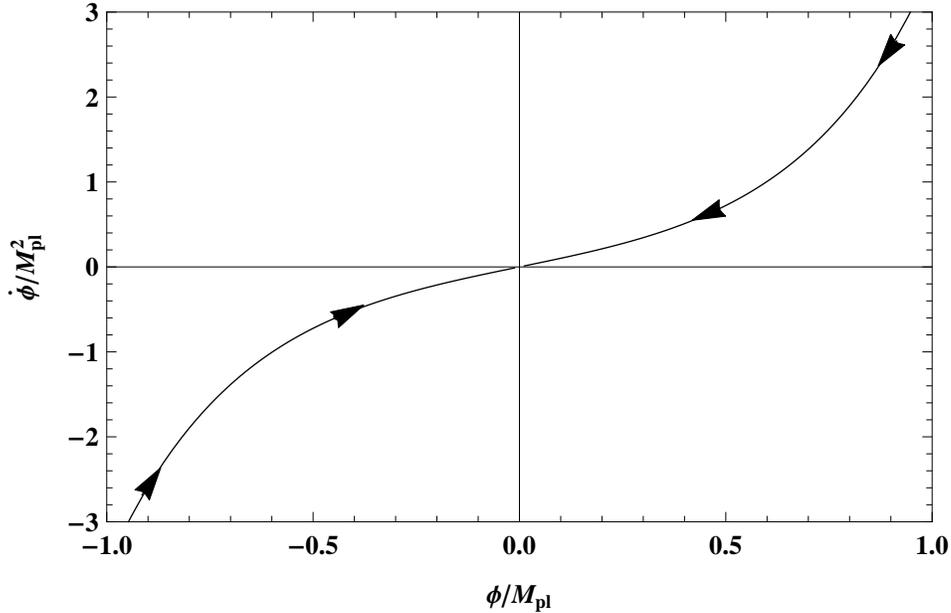}}
\caption{Phase space structure of the solution (\ref{eq33}).  In this numerical simulation we
have used:  $\frac{\phi_0}{M_{\text{pl}}}=5$ (right top plot), $\frac{\phi_0}{M_{\text{pl}}}=-5$ (left down plot) and  $\gamma=-10$. \label{fig5}}
\end{figure}

\begin{figure}[t]
\includegraphics[width=1\linewidth,scale=1]{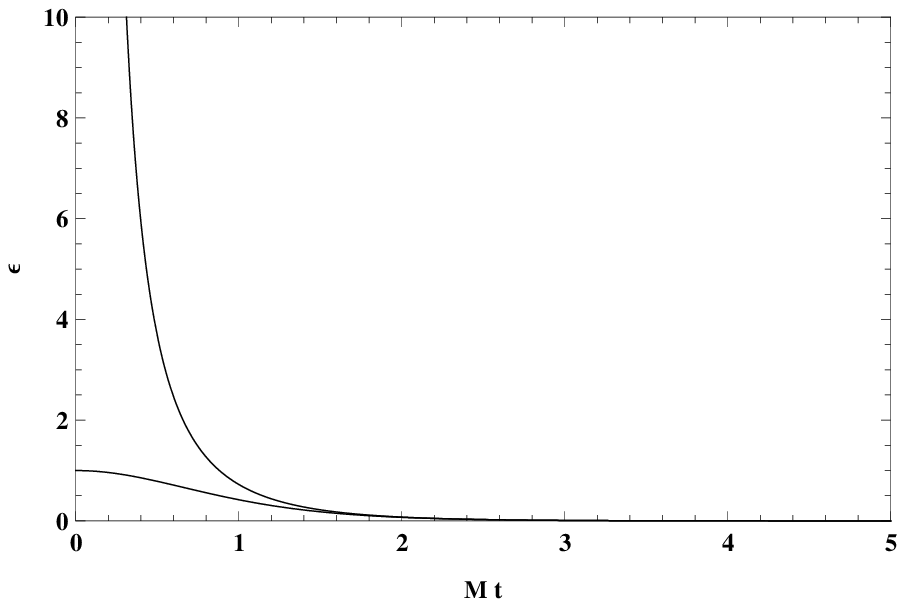}
\caption{Evolution of the first slow-roll parameter $\epsilon\equiv -\dot{H}/H^2$. (top) Solution (\ref{eq27}) and  (bottom) solution (\ref{eq33}).  \label{fig6}}
\end{figure}

\noindent Now we proceed to verify whether the solutions given by Eqs. (\ref{eq27}) and (\ref{eq33}) are attractors solutions or not (the same formalism can be used for the other solutions).
In this ways,  we numerically solved Eqs. (\ref{eq8})-(\ref{eq10}) under the assumption (\ref{eq11}) and also,
we used  the potentials $V(\phi)$ found for each case. In this sense, the first case studied was the solution given by Eq. (\ref{eq27}).
Additionally,  various initial conditions were considered for it (for simplicity, only four
choices are shown). Thereby, we obtained the phase space diagram shown in Fig. \ref{fig4}, in which,  we see that the phase space flow converge in distinct points depending on initial conditions. This observed pattern is related to the form of the potential  (\ref{eq28}), since it is a periodic function of $\phi$ and it preserves its form under the translation $\phi\rightarrow \phi+\text{const.}$ (a similar result was obtained in Ref.~\refcite{motohashi1}, for $\alpha=0$). Besides, an analogous behavior it is obtained using other initial conditions and suitable values for the model parameters. For the solution (\ref{eq33}), as $\phi$ and $\dot{\phi}$ are not defined in $t=0$, it's necessary to recover the integration constant in the analytic solution given by Eq. (\ref{eq35}), 
(i.e. $t\rightarrow t+\frac{\sqrt{2(1+2\gamma)}}{2M M_{\text{pl}}}C$), and  determine it according to an arbitrary initial condition. 
So, in Fig. \ref{fig5} we display the phase space diagram  associated to the solution given by Eq. (\ref{eq33}), in which we see a typical attractor behavior where the trajectories converges to $\phi=0$ and $\dot{\phi}=0$. This implies that the inflaton approaches to the global minimum of the potential at $\phi=0$ (see Fig. \ref{fig3}). Furthermore, for various initial conditions and for small values of the variables $(\phi,\dot{\phi})$,  all trajectories  overlap. The trajectories are separated for larger  values of the variables. By last, the inflationary solutions studied above, fulfill  the first slow-roll condition $|\epsilon|\ll 1$ (see Fig. \ref{fig6}).

\section{Conclusions}\label{sec5}
\noindent In this work,  we have studied a scalar-tensor model of inflation in which the action for the scalar field has the  kinetic term non-minimally coupled to Einstein tensor (NMDC). In this context, instead of using the usual slow-roll approximation to analyze the inflation dynamics, we have used the constant-roll condition given by Eq. (\ref{eq11}), also, using the Hamilton-Jacobi-like formalism, an ansatz for the Hubble parameter (see Eq. (\ref{eq18})) and some restrictions on the model parameters, we found new exact solutions for the inflaton potential, which include power-law, de Sitter, quadratic hilltop  and natural inflation, among others. For natural inflation, is necessary that the restriction $\gamma>-\frac{1}{2}$ is satisfied, and for hilltop inflation the restriction is $\gamma>-\frac{1}{6}$. The restriction for the potential given by  Eq. (\ref{eq34}) is $\gamma<-\frac{1}{2}$. Also, a phase space analysis was performed and it was shown that the  exact solutions given by  Eqs. (\ref{eq27}) and (\ref{eq33}) are attractors  (see Figs. \ref{fig4} and \ref{fig5} ). Also, these inflationary solutions fulfill  the slow-roll condition  $|\epsilon|\ll 1$ (see Fig. \ref{fig6}). Aside from this, to decide if  the phenomenological inflationary solutions found in this work are viable or not,  it's necessary to investigate in detail the  most recent observational constraints on model parameters (like those studied in Refs.~\refcite{motohashi2} and \refcite{shinji}) and also, a full  analysis of the evolution of scalar and tensor perturbations like that studied in Ref.~\refcite{motohashi1} must be performed. But that kind of analysis is beyond the scope of the present work and could be addressed later. Finally, we must emphasize that the main goal of the present work was to find new exact constant-roll inflationary solutions derived from the ansatz (\ref{eq18}), therefore,  it's possible that  
other solutions may exist for this model.

\end{document}